\documentclass[aps,pra,superscriptaddress,showpacs,twocolumn]{revtex4-1}
\usepackage{bm}
\usepackage{amsmath}
\usepackage{float}
\usepackage{longtable}
\usepackage{dcolumn}
\usepackage{nicefrac}
\newcolumntype{.}{D{x}{}{-1}}

\newcolumntype{w}[1]{D{.}{.}{#1}}
\newcommand{\lbr}{\langle}
\newcommand{\rbr}{\rangle}

\usepackage{graphicx}
\usepackage{xcolor}
\allowdisplaybreaks

\begin{document}

\title{Relativistic corrections to the Bethe logarithm for the $\bm{2\,^3\!S}$ and $\bm{2\,^3\!P}$ states of He}

\author{Vladimir A. Yerokhin}
\affiliation{Center for Advanced Studies, Peter the Great St.~Petersburg Polytechnic University,
  Polytekhnicheskaya 29, 195251 St.~Petersburg, Russia}

\author{Vojt\v{e}ch Patk\'o\v{s}}
\affiliation{Faculty of Mathematics and Physics, Charles University,  Ke Karlovu 3, 121 16 Prague
2, Czech Republic}

\author{Krzysztof Pachucki}
\affiliation{Faculty of Physics, University of Warsaw,
             Pasteura 5, 02-093 Warsaw, Poland}

\begin{abstract}

With this work we start a project calculating the QED contribution of order $\alpha^7m$ to the
$2\,^3\!P$--$2\,^3\!S$ transition energy in helium, aiming for an accurate determination of the
nuclear charge radius $r_E$ from measurements of the corresponding transition frequency. Together
with the complementary determination of $r_E$ from muonic  helium, this project will provide a
stringent test of universality of electromagnetic  interactions of leptons in the Standard Model. We
report a calculation of the relativistic  corrections to the Bethe logarithm for the $2\,^3\!S$
and $2\,^3\!P$ states, which is the most numerically demanding part of the project.

\end{abstract}

\maketitle

\section{Introduction}
One of the prominent low-energy  tests of the Standard Model (SM) with a possible signature of new
physics is based on a comparison of the Lamb shift in muonic hydrogen $\mu$H and electronic
hydrogen H. The lepton universality of SM implies that the same physical laws and physical
constants define the energy levels in H and $\mu$H. However, it has been found that the
proton root-mean-square charge radius, extracted from the comparison of theory and experiment for
the Lamb shift, turned out to be significantly different for the electronic \cite{mohr:16:codata}
and muonic \cite{pohl:10,antognini:13} hydrogen,
\begin{align*}
r_p({\rm H}) &= \,0.875\,9\,(77)\,\ \text{fm}\,,\\
r_p({\mu \rm H}) &=\, 0.840\,87\,(39)\,\ \text{fm}.
\end{align*}
This $4.5\,\sigma$ discrepancy, known as the proton radius puzzle, may signal the existence of
interactions that are not accounted for in the Standard Model. Several experiments aiming to
resolve the puzzle have been accomplished recently, namely the measurement of the $2S$--$4P$
transition energy in Garching \cite{beyer:17}, $2S$--$2P$ Lamb shift in Toronto \cite{hessels:18},
and two measurements of the $1S$--$3S$ transition energy performed in Paris \cite{fleurbaey:18} and
in Garching \cite{pohl:talk}. These experiments yield conflicting results for the proton charge
radius, which does not solve the puzzle but suggests the presence of unknown systematic effects in
hydrogen measurements. Further experiments directed to clarify the proton radius puzzle are
currently being pursued, notably, measurements of the $1S$--$2S$ transition energy in He$^+$
\cite{eikema:priv,udem:priv}, transitions between circular Rydberg states in H-like ions
\cite{tan:11}, and the direct comparison of the cross sections of the $e$-$p$ versus $\mu$-$p$
elastic scattering \cite{gilman:13}.

An alternative way to solve the proton charge radius puzzle can be gained through spectroscopy
of the helium atom. Specifically, a comparison of the nuclear charge radius from the helium
spectroscopy with the radius from muonic helium, expected soon from the CREMA collaboration
\cite{pohl:16:leap}, would provide an independent test of the lepton universality in atomic
systems. On the experimental side, several transition energies in the helium atom are already known
with an accuracy sufficient for determining the nuclear charge radius on a $10^{-3}$ fractional
level \cite{florence:94,rooij:11,luo:13,notermans:14,luo:16,shuiming:2017,rengelink:18}. In order to
achieve a similar level of accuracy in theoretical predictions, one needs to improve the previous
helium calculations \cite{pachucki:06:hesinglet,yerokhin:10:helike} by completing the next-order
term of the NRQED expansion, namely the $\alpha^7 m$ correction. This is a very challenging
theoretical problem. Among few-electron atoms, it has so far been solved only for the helium fine
structure \cite{pachucki:06:prl:he,pachucki:09:hefs,pachucki:10:hefs}.

With this paper, we start a project calculating the complete $\alpha^7 m$ correction for
energy levels of two-electron atoms. At present, we restrict ourselves to the triplet states, for
which the nonrelativistic wave function $\phi(\vec r_1, \vec r_2)$ vanishes at $\vec r_1 = \vec r_2$.
As a result, the whole class of the so-called contact operators does not contribute, thus
making the derivation of $\alpha^7\,m$ operators more tractable. An improved theory of the triplet
states will allow the determination of the nuclear charge radius from the $2^3\!S$--$2^3\!P$
transition in $^4$He, which was accurately measured by the Hefei group \cite{shuiming:2017}.

The present status of theory of helium energy levels complete up to order $\alpha^6 m$ is described
in our recent review \cite{pachucki:17}. The next-order $\alpha^7m$
contribution can be represented as a sum of three parts,
\begin{equation} \label{eq1}
E^{(7)} = \Bigl< H^{(7)} \Bigr> + 2\,\biggl\langle H^{(4)}\,\frac{1}{(E - H)'}\,H^{(5)}\biggr\rangle + E_L\,,
\end{equation}
where $H^{(4)}$, $H^{(5)}$, and $H^{(7)}$ are the effective Hamiltonians of order $\alpha^4m$,
$\alpha^5m$, and $\alpha^7m$, respectively; $H$ and $E$ are the nonrelativistic Hamiltonian and its
eigenvalue, respectively; and $E_L$ is the low-energy contribution, also known as the relativistic
correction to the Bethe logarithm, which is the main subject of this work.

Out of the three terms contributing to $E^{(7)}$, the relativistic correction to the Bethe
logarithm is numerically the most demanding one and thus is the crucial part of the whole
$\alpha^7\,m$ project. The calculation of (the spin-dependent part of) such a correction was first
performed in Ref.~\cite{pachucki:00:jpb} for the fine structure of helium and later improved in
Refs.~\cite{pachucki:09:hefs,pachucki:10:hefs}. For the Coulomb two-center systems (H$_2^+$,
HD$^+$, etc.), the relativistic corrections to the Bethe logarithm were calculated by Korobov {\em
et al.}~\cite{korobov:13}. The goal of the present work is to calculate the spin-independent
low-energy correction $E_L$ for the $2\,^3\!S$ and $2\,^3\!P$ states of helium.

\section{Nonrelativistic low-energy contribution}

The leading nonrelativistic (dipole) low-energy contribution of order $\alpha^5\,m$ is given by
\begin{eqnarray}
E_{L0}(\Lambda) &=& \frac{e^2}{m^2}\int_{k<\Lambda} \frac{d^3 k}{(2\,\pi)^3\,2\,k}\,
\left(\delta^{ij}-\frac{k^i\,k^j}{k^2}\right)\,
\nonumber \\ &&\times
\left< P^i\,\frac{1}{E-H-k}\,P^j \right>\,,
\label{05}
\end{eqnarray}
where $\vec P = \vec p_1 + \vec p_2$ and $\Lambda$ is the high-momentum cutoff. $E_{L0}(\Lambda)$
diverges when $\Lambda \to \infty$ and requires subtraction of the leading terms in
the large-$\Lambda$ asymptotics.

Performing the angular integration  and dropping the overall prefactor $\alpha^5 m$, one obtains
\begin{equation}
  E_{L0}(\Lambda) = \frac{2}{3\,\pi}\int_0^\Lambda k\,dk\,P(k)\,,
\end{equation}
where
\begin{equation}
P(k) = \left<  \vec{P}\,\frac{1}{E-H-k}\,\vec{P} \right>\,.
\end{equation}
The large-$k$ expansion of $P(k)$ is
\begin{equation}
  k\,P(k) = - \langle P^2\rangle +\frac{D}{k} + \ldots\,,
\end{equation}
where $D=2\,\pi\,Z\,\langle\delta^3(r_1) + \delta^3(r_2)\rangle$. The finite part of the low-energy
contribution is then defined by dropping terms proportional to $\Lambda$ and $\ln(2\,\Lambda)$
as
\begin{equation}
E_{L0} = -\frac{2}{3\,\pi}\,D\,\ln k_0\,,
\end{equation}
where $\ln k_0$ is the standard Bethe logarithm,
\begin{align}
  \ln k_0 =&\  -\frac{1}{D}\int_0^\infty dk\biggl[
    k\,P(k) +\langle P^2\rangle -\frac{D}{k}\,\theta(2k-1)\biggr]
    \nonumber \\
  =&\
  \frac{\big \langle \vec P \,(H-E)\,\ln \big[2\,(H-E)\big]\,\vec P \big\rangle}
       {2\,\pi\,Z\,\big\langle\sum_a \delta^3(r_{\rm a})\big\rangle}\,,
\end{align}
and $\theta(x)$ is the Heaviside step function.

The numerical calculation of the Bethe logarithm for the helium atom remained for a long time a
very difficult problem \cite{schwartz:61,baker:93}, which has been successfully solved only
relatively recently \cite{korobov:99,drake:99:cjp,korobov:04}.

\section{Relativistic low-energy corrections}
\label{sec:rel}

There are three types of relativistic corrections of order $\alpha^7 m$ to the low-energy
contribution (\ref{05}),
\begin{equation}
  E_L = E_{L1} + E_{L2} + E_{L3}\,.
\end{equation}
The first part $E_{L1}$ is a perturbation of the nonrelativistic low-energy contribution $E_{L0}$
in Eq.~(\ref{05}) by the Breit Hamiltonian $H^{(4)}$, the second part $E_{L2}$ is induced by the
relativistic correction to the current operator $\vec P$, whereas the third term $E_{L3}$ is the
retardation correction. All of these corrections are defined as  remainders after dropping
divergent in $\Lambda$ terms, such as $\Lambda$ and $\ln\Lambda$.

\subsection{Breit correction $\bm{E}_{\bm{L1}}$}

The low-energy contribution perturbed by the (spin-independent part of the) Breit Hamiltonian
$H^{(4)}$ is,
\begin{align}
E_{L1}(\Lambda) &\ = \frac{2}{3\,\pi} \int_0^{\Lambda} kdk\,P_{L1}(k)\,,
\end{align}
where
\begin{align}
P_{L1}(k) &\ = 2 \left< H^{(4)} \frac1{(E-H)'}\, \vec{P}\,  \frac1{E-H-k} \, \vec{P} \right>
 \nonumber \\ \label{eqL12}
&\ +  \left< \vec{P}\,  \frac1{E-H-k} \,\Bigl[ H^{(4)} - \lbr H^{(4)}\rbr\Bigr]  \frac1{E-H-k} \,\vec{P} \right> \,,
\end{align}
where (with $r\equiv r_{12}$)
\begin{align}\label{H4}
H^{(4)} = &\  -\frac{1}{8}\,\bigl(p_1^4+p_2^4\bigr)+
\frac{Z\,\pi}{2}\,\Bigl[\delta^3(r_1)+\delta^3(r_2)\Bigr]+\pi\,\delta^3(r)
\nonumber\\&
-\frac{1}{2}\,p_1^i\,
\biggl(\frac{\delta^{ij}}{r}+\frac{r^i\,r^j}{r^3}\biggr)\,p_2^j\,.
\end{align}

The large-$k$ expansion of $P_{L1}(k)$ is given by
\begin{align} \label{eq:PL1}
k P_{L1}(k) &\ = A_1 + \frac{B_1}{\sqrt{k}} + \frac{C_1\,\ln k}{k} + \frac{D_1}{k} + \ldots\,,
\end{align}
where the asymptotic constants are derived in Appendix~\ref{sec:app1}.

\begin{widetext}
We construct the finite part of the Breit correction as
\begin{align}
E_{L1} = &\ \frac{2}{3\,\pi} \int_0^{\infty} dk\,\biggl[k P_{L1}(k)
- A_1 - \frac{B_1}{\sqrt{k}}
- \biggl(\frac{C_1\,\ln k}{k} + \frac{D_1}{k}\biggr) \theta(k-1)\biggr]
 \\ = &\
 \frac{2}{3\pi} \biggl\{ \int_0^K kdk\,P_{L1}(k)
+ \int_K^{\infty} dk\,
\biggl[ k P_{L1}(k)
- A_1  - \frac{B_1}{\sqrt{k}}  - \frac{C_1\ln k}{k}  - \frac{D_1}{k}\biggr]
\nonumber \\ &\
 - \biggl[ A_1\,K + 2\,B_1\,\sqrt{K} + \frac{C_1}{2}\ln^2K + D_1\,\ln K\biggr]\biggr\}\,,
\label{eqL1}
\end{align}
where $K \ge 1$ is a free parameter.
\end{widetext}

\subsection{Current correction $\bm{E}_{\bm{L2}}$}

The second low-energy contribution of order $\alpha^7m$ is induced by a correction to the current
operator in Eq.~(\ref{05}), $\vec{P} \to \vec{P} + \delta \vec{j}$,  with
\begin{align} \label{eq:deltaj}
  \delta j^i =\ & i\Bigl[H^{(4)}, r_1^i + r_2^i\Bigr]\nonumber \\ =\ &
  -\frac{1}{2}\,\left(p_1^i\,p_1^2+p_2^i\,p_2^2\right) -\frac{1}{2}
   \left(\frac{\delta^{ij}}{r} + \frac{r^ir^j}{r^3}\right)\, \left(p_1^j+p_2^j\right)\,,
\end{align}
where $\left[\,\,,\, \right]$ denotes a commutator. The corresponding low-energy correction is
\begin{align}
  E_{L2}(\Lambda) &\ = \frac{4}{3\,\pi} \int_0^{\Lambda} kdk\,P_{L2}(k)\,,
\end{align}
where
\begin{align} \label{eq:PL2:0}
P_{L2}(k) =  \left< \vec{\delta j} \, \frac1{E-H-k} \, \vec{P} \right> \,.
\end{align}
The large-$k$ expansion of $P_{L2}(k)$ has the same form as that of $P_{L1}(k)$,
\begin{align} \label{eq:PL2}
k\,P_{L2}(k) &\ = A_2 + \frac{B_2}{\sqrt{k}} + \frac{C_2\,\ln k}{k} + \frac{D_2}{k} + \ldots\,,
\end{align}
where the asymptotic constants are derived in Appendix~\ref{sec:app2}.

\begin{widetext}
The finite part of the low-energy correction is constructed as
\begin{align}
E_{L2} = &\ \frac{4}{3\,\pi} \int_0^{\infty} dk\,\biggl[k\,P_{L2}(k)
- A_2 - \frac{B_2}{\sqrt{k}}
- \biggl(\frac{C_2\,\ln k}{k} + \frac{D_2}{k}\biggr) \theta(k-1)\biggr]
   \\
      = &\ \frac{4}{3\pi} \biggl\{ \int_0^K kdk\,P_{L2}(k) + \int_K^{\infty} kdk\,
      \biggl[ P_{L2}(k)
        - \frac{A_2}{k} - \frac{B_2}{k^{3/2}}  - \frac{C_2\ln k}{k^2}  - \frac{D_2}{k^2}\biggr]
 \nonumber \\ &
 - \biggl[A_2\,K + 2B_2\sqrt{K} + \frac{C_2}{2}\ln^2K + D_2\ln K\biggr]\biggr\}\,,
\label{eqL2}
\end{align}
where $K \ge 1$ is a free parameter.

\subsection{Retardation correction $\bm{E}_{\bm{L3}}$}

A retardation correction to the low energy contribution is
\begin{align}
E_{L3}(\Lambda) &\ = \frac{2}{3\,\pi} \int_0^{\Lambda} kdk\,P_{L3}(k)\,,
\end{align}
where
\begin{align} \label{eq:PL3:0}
P_{L3}(k) = &\ \frac{3}{8\pi} \int d\Omega_k
 \left( \delta^{ij} - \frac{k^i\,k^j}{k^2}\right)\,\sum_{a,b=1,2}
\,\delta_{k^2}\,\biggl\langle p_a^i\,e^{i\,\vec k\cdot\vec r_a}\, \frac1{E-H-k} \,p_b^j\,e^{-i\,\vec k\cdot\vec r_b} \biggr\rangle\,,
\end{align}
where $\delta_{k^2}\langle \ldots \rangle$ denotes the quadratic in $k$ term of the small-$k$
expansion of the exponential functions in the matrix element $\langle \ldots \rangle$. Performing
the expansion and integrating over angular variables, we obtain
\begin{align} \label{eq:PL3b}
P_{L3}(k) = &\ \frac{3}{8\pi} \int \Omega_k
 \left( \delta^{ij} - \frac{k^i\,k^j}{k^2}\right)\,\sum_{a,b=1,2}\biggl\langle p_a^i\,(\vec k\cdot\vec r_a)
\,\frac1{E-H-k} \,(\vec k\cdot\vec r_b)\, p_b^j
-  p_a^i\, (\vec k\cdot\vec r_a)^2\, \frac1{E-H-k} \, p_b^j \biggr\rangle
 \\
= &\ \frac{k^2}{10} \Biggl[
3 \left< \bigl( p_1^{i} r_1^{j} + p_2^{i} r_2^{j}\bigr)^{(2)} \, \frac1{E-H-k} \,
         \bigl( r_1^{j} p_1^{i} + r_2^{j} p_2^{i}\bigr)^{(2)} \right>
  -\frac{5}{2\,k}\, \left< \vec L^2 \right>
         \nonumber \\ &\
  -2\, \left< \biggl[ p_1^i \bigl( 2\,\delta^{ij}r_1^2 - r_1^ir_1^j\bigr)
                    +p_2^i \bigl( 2\,\delta^{ij}r_2^2 - r_2^ir_2^j\bigr) \biggr] \, \frac1{E-H-k} \,
         \bigl( p_1^j + p_2^j\bigr) \right>\Biggr]\,,
         \label{eq:PL3c}
\end{align}
where $(a^ib^j)^{(2)} = (a^ib^j + a^jb^i)/2 - \vec a\cdot\vec b\,\delta^{ij}/3$
and $\vec L = \vec r_1\times\vec p_1 + \vec
r_2\times\vec p_2$. The large-$k$ expansion of $P_{L3}(k)$ is of the form
\begin{align} \label{eq:PL3}
k\,P_{L3}(k) &\ = G_3\,k^2 + F_3\,k + A_3 + \frac{B_3}{\sqrt{k}} + \frac{C_3\,\ln k}{k} + \frac{D_3}{k} + \ldots\,,
\end{align}
where the asymptotic constants are derived in Appendix~\ref{sec:app3}.

We construct the finite part of the retardation correction as
\begin{align}
E_{L3} = &\ \frac{2}{3\,\pi} \int_0^{\infty} dk\,\Biggl\{k\,P_{L3}(k) - k^2\,G_3 - k\,F_3
- A_3 - \frac{B_3}{\sqrt{k}}
- \biggl[\frac{C_3\,\ln k}{k} + \frac{D_3}{k}\biggr] \theta(k-1)\biggr\}\,
  \\
  =&\  \frac{2}{3\pi} \biggl\{ \int_0^K dk\,k P_{L3}(k) + \int_K^{\infty} dk\,
\biggl[ k P_{L3}(k)  - G_3 k^2
- F_3 k - A_3 - \frac{B_3}{\sqrt{k}}  - \frac{C_3\ln k}{k}  - \frac{D_3}{k}\biggr]
\nonumber \\ &
- \biggl[ G_3 \frac{K^3}{3}
+ F_3 \frac{K^2}{2} + A_3\,K + 2\,B_3\,\sqrt{K} +\frac{C_3}{2}\ln^2K + D_3\ln K\biggr] \Biggr\}\,,
\label{eqL3}
\end{align}
where $K \ge 1$ is a free parameter.
\end{widetext}

%%%%%%%%%%%%%%%%%%%%%%%%%%%%%%%%%%%%%%%%%%%%%%%%%%%%%%%%%%%%%%%%%%%%%%%%%%%%%%%%%

\section{Numerical evaluation}

\subsection{Transformation to a regularized form}

The Breit Hamiltonian $H^{(4)}$ [Eq.~(\ref{H4})] contains singular operators ($\delta(r_a)$,
$p_a^4$) which complicates numerical evaluations of the Breit correction $E_{L1}$. In order to
achieve high numerical accuracy, we transform Eq.~(\ref{eqL12}) to a more
regular form. Specifically, by using the identity
\begin{align} \label{eq:HA}
  H^{(4)}|\phi\rbr =& H_A^{(4)}|\phi\rbr + \bigl\{ H-E,Q \bigr\}|\phi\rbr\,,
\end{align}
where $|\phi\rbr$ is the eigenfunction $H$ with energy $E$,
\begin{align}
Q =& - \frac14 \left( \frac{Z}{r_1} + \frac{Z}{r_2} - \frac{2}{r} \right)
\end{align}
and
\begin{align}
H_A^{(4)} |\phi\rangle= &  \biggl[-\frac12 (E-V)^2 + \frac14 \nabla_1^2 \nabla_2^2 - \frac{Z}{4} \frac{\vec{r}_1}{r_1^3}\cdot\vec{\nabla}_1
\nonumber \\ &
- \frac{Z}{4} \frac{\vec{r}_2}{r_2^3}\cdot\vec{\nabla}_2
-\frac{1}{2}\,p_1^i\,
\biggl(\frac{\delta^{ij}}{r}+\frac{r^i\,r^j}{r^3}\biggr)\,p_2^j\biggr]|\phi\rangle\,,
\end{align}
and with $V = -Z/r_1-Z/r_2+1/r$, we transform the first term in the right-hand-side of
Eq.~(\ref{eqL12}) to a regularized form,
\begin{align}
P_{\rm pwf}(k) = &\ 2 \left< H_A^{(4)} \frac1{(E-H)'}\, \vec{P}\,  \frac1{E-H-k} \, \vec{P} \right>
 \nonumber \\ &
 - 2 \left< \bigl[ Q - \lbr Q\rbr \bigr] \, \vec{P}\,  \frac1{E-H-k} \, \vec{P} \right>\,.
\end{align}

Furthermore, by using the identity
\begin{align}
H^{(4)} = H^{(4)}_B + \bigl\{ H-E,Q_B \bigr\} -\frac{1}{2}\,(H-E)^2\,,
\end{align}
where
\begin{align}
Q_B = -\frac{E}{2}  - \frac14 \left( \frac{Z}{r_1} + \frac{Z}{r_2} - \frac{2}{r} \right)\,,
\end{align}
and
\begin{widetext}
\begin{align}
  H^{(4)}_B &\ = -\frac12 (E-V)\,\left(E-\frac{1}{r}\right) + \frac14 \nabla_1^2 \nabla_2^2
  - \frac{Z}{4} \vec p_1\left(\frac{1}{r_1}+\frac{1}{r_2}\right)\vec p_1
  - \frac{Z}{4} \vec p_2\left(\frac{1}{r_1}+\frac{1}{r_2}\right)\vec p_2
-\frac{1}{2}\,p_1^i\,\biggl(\frac{\delta^{ij}}{r}+\frac{r^i\,r^j}{r^3}\biggr)\,p_2^j\,,
\end{align}
we transform the second term in the right-hand-side of Eq.~(\ref{eqL12}) to the regularized form,
\begin{align}
P_{\rm ver}(k) &\ =  \left< \vec{P}\,  \frac1{E-H-k} \,\Bigl[ H^{(4)}_B -2 k Q_B -\frac{k^2}{2}- \Bigl< H^{(4)}_B\Bigr>\Bigr]  \frac1{E-H-k} \,\vec{P} \right>
%\nonumber \\ &
  - \left<\bigl[ 2\vec{P}\,Q_B+k\vec{P}\bigr]  \frac1{E-H-k}  \,\vec{P} \right>  - \frac12 \left< \vec{P}^2 \right>\,.
\end{align}
\end{widetext}

\subsection{Angular decomposition}

In our approach, we express all wave functions and perform the angular momentum algebra in
Cartesian coordinates. The reference-state wave functions of the $^3\!S$ and $^3\!P^o$ symmetry are
represented as
\begin{align}
\phi\left(^{3}\!S\right) &\ = F(r_1,r_2,r) - (1\leftrightarrow2)\,, \\
\phi^i\left(^{3}\!P^o\right) &\ = r^i_1 \, F(r_1,r_2,r) - (1\leftrightarrow2)\,,
\end{align}
where the scalar functions $F$ are linear combinations of exponential functions,
\begin{align} \label{wf}
F(r_1,r_2,r) = \sum_i c_i \exp\left(-\alpha_i r_1 - \beta_i r_2 - \gamma_i r \right)\,.
\end{align}
The wave functions are normalized by $\langle\phi|\phi\rangle = 1$ and $\langle\phi^i|\phi^i\rangle
= 1$.

The angular decomposition of formulas in Sec.~\ref{sec:rel} is mostly performed in the same way as for
the nonrelativistic Bethe logarithm. In that case, for the $^3\!S$ reference state, only $^3P^o$
intermediate states contribute,
\begin{align}
P(k) &\ = \left< \phi\left(^{3}\!S\right) \Bigl|\, P^i \,  \left(\frac1{E-H-k}\right)_{^3P^o}  \, P^i  \, \Bigr|\phi \left(^{3}\!S\right)\right>
 \nonumber \\
     &\ = \left< \phi\left(^{3}\!S\right) \Bigl|\, P^i \Bigr| \, \delta \phi^i \left(^{3}\!P^o\right)\right>\,,
\end{align}
where $\delta \phi^i$ is the perturbed wave function of the $^{3}\!P^o$ symmetry,
\begin{align}
     \delta \phi^i \left(^{3}\!P^o\right) =
        \sum _n \frac{\phi^i_n \left(^{3}\!P^o\right)}{E-E_n-k}
        \Bigl<\phi^k_n \left(^{3}\!P^o\right) \Bigl| \, P^k \, \Bigr| \phi \left(^{3}\!S\right)\Bigr>\,.
\end{align}

The angular decomposition for the $^3P^o$ reference state is performed by using the identity
\begin{align}
j^i \phi^k = &\ \frac13\,\delta^{ik} \vec{j}\cdot\vec{\phi} + \frac12\, \epsilon_{ikl} \bigl(
\vec{j}\times\vec{\phi}\bigr)_l
 \nonumber \\ &
+ \frac12\biggl(j^i \phi^k + j^k \phi^i -\frac23 \,\delta^{ik}\, \vec{j}\cdot\vec{\phi}\biggr)\,.
 \end{align}
The three terms in the right-hand-side of the above expression give rise to contributions from the
$^3S$, $^3\!P^e$, and $^3\!D^e$ intermediate states, respectively,
\begin{align}
P(k) = &\  P_{L=0}(k) + P_{L=1}(k) + P_{L=2}(k)
 \nonumber \\ = &\
 \frac13\, \left< \Psi_0 \, \Bigl| \left(\frac{1}{E-H-k}\right)_{^3S} \Bigr| \Psi_0 \right>
 \nonumber \\ &
+ \frac12\, \left< {\Psi}^i_1  \, \Bigl| \left( \frac{1}{E-H-k}\right)_{^3\!P^e}\Bigr| \Psi^i_1 \right>
 \nonumber \\ &
+ \frac14\, \left< {\Psi}^{ik}_2  \, \Bigl| \left( \frac{1}{E-H-k}\right)_{^3\!D^e} \Bigr| \Psi^{ik}_2 \right>
\,,
\end{align}
where $\Psi_0 = \vec{j}\cdot \vec{\phi}$, $\vec{{\Psi}}_1 =  \vec{ j}\times\vec{\phi}$, and
${\Psi}_2^{ik} = j^i \phi^k + j^k \phi^i -\frac23 \, \delta^{ik} \bigl(
\vec{j}\cdot\vec{\phi}\bigr)$.

A more complicated situation arises in the evaluation of the symmetric part of the $E_{L3}$
contribution for the $^3\!P$ reference state [the first term in brackets in Eq.~(\ref{eq:PL3c}),
$P_{L3}^{\rm sym}$]. In order to perform the angular decomposition in this case, we use the following
identity,
\begin{align} \label{30}
    \frac12\sum_{a=1,2}\bigl( r_a^i p_a^j + r_a^jp_a^i\bigr)\,\phi^k
    =&\ T^{ijk} + \epsilon^{ikl}\,T^{lj} + \epsilon^{jkl}\,T^{li}
 \nonumber \\ &
    +\delta^{ik}\,T^j + \delta^{jk}\,T^i + \delta^{ij}\,T'^k\,,
\end{align}
where $T^{ijk}$, $T^{ij}$, and $T^i$ are the components of the (symmetric and traceless)
irreducible tensors of the first, second, and third rank, respectively,
\begin{align}
    T^{ijk} =& \ \sum_a (r_a^i\,p_a^j\,\phi^k)^{(3)}\,,\\
    T^{ij} =& \frac1{12}\,\sum_a \Bigl[
    \epsilon^{jlm}\,\bigl(r_a^{i}\,p_a^{l}+r_a^{l}\,p_a^{i}\bigr)\,\phi^m
  \nonumber \\ &
  + \epsilon^{ilm}\,\bigl(r_a^{j}\,p_a^{l}+r_a^{l}\,p_a^{j}\bigr)\,\phi^m
    \Bigr] \,, \\
    T^{i} =&  \frac1{20}\,\sum_a \Bigl[
    3\,\bigl(r_a^{i}\,p_a^{l}+r_a^{l}\,p_a^{i}\bigr)\,\phi^l - 2\,r_a^l\,p_a^l\,\phi^i
    \Bigr]\,, \\
    T'^{i} =& \frac{1}{10} \sum_a \Bigl[ 4 r_a^l\,p_a^l\,\phi^i - r_a^{i}\,p_a^{l}\,\phi^l
    -  r_a^{l}\,p_a^{i}\,\phi^l\Bigr]\,.
\end{align}
Using this identity, we express $P_{L3}^{\rm sym}$ as a sum of the $L = 1$, $L = 2$, and $L = 3$
parts,
\begin{equation}
  P_{L3}^{\rm sym}(k) = P_{L3,1}^{\rm sym}(k) + P_{L3,2}^{\rm sym}(k) + P_{L3,3}^{\rm sym}(k)\,,
\end{equation}
where
\begin{align}
P_{L3,1}^{\rm sym}(k) =& \ \frac{3\,k^2}{2}\,\frac{4}{3}\,\Bigl\langle  T^{i}\Big|\left(\frac{1}{E-H-k}\right)_{^3P^o}\Bigr|T^i\Bigr\rangle\,,\\
P_{L3,2}^{\rm sym}(k) =& \  \frac{3\,k^2}{2}\,\frac{6}{5}\,\Bigl\langle  T^{ij}\Big|\left(\frac{1}{E-H-k}\right)_{^3D^o}\Bigr|T^{ij}\Bigr\rangle\,,\\
P_{L3,3}^{\rm sym}(k) =& \ \frac{3\,k^2}{2}\,\frac{1}{5}\,\Bigl\langle  T^{ijk}\Big|\left(\frac{1}{E-H-k}\right)_{^3F^o}\Bigr|T^{ijk}\Bigr\rangle\,.
\end{align}

Wave functions of the different symmetries in Cartesian coordinates required in this work are
summarised in Appendix~\ref{app:wf}.

\subsection{Numerical details}

Numerical evaluation of the relativistic corrections to the Bethe logarithm was performed according
to Eqs.~(\ref{eqL1}), (\ref{eqL2}), and (\ref{eqL3}). The general scheme of the computation was
similar to the one developed in our previous calculation of the helium fine structure
\cite{pachucki:09:hefs} (as described in Sec. V.E of that work). Numerical cancelations, however,
were much larger in the present work, because of a greater number of asymptotic expansion terms
that needed to be separated out.

The low-energy part of the $k$ integral, $k \in (0,K)$ with $K = 10$--$100$, was evaluated
analytically after diagonalizing the matrix representation of the Schr\"odinger Hamiltonian. In
order to perform the high-energy part of the integral, $k \in (K,\infty)$, we calculated the
integrand for several hundreds different values of $k \in (5,10\,000)$, subtracted the
contributions of the known asymptotic expansion coefficients, fitted the residual, and calculated
the integral analytically. For fitting of the subtracted integrands $w_{Li}$,
\begin{align}
w_{L1}(k) = &\ k\, P_{L1}(k) - A_1 - \frac{B_1}{\sqrt{k}}  - \frac{C_1\ln k}{k}  - \frac{D_1}{k} \,, \\
w_{L2}(k) = &\ k\, P_{L2}(k) - A_2 - \frac{B_2}{\sqrt{k}}  - \frac{C_2\ln k}{k}  - \frac{D_2}{k} \,, \\
w_{L3}(k) = &\ k\, P_{L3}(k)  - G_3 k^2
- F_3 k
 \nonumber \\ &
- A_3 - \frac{B_3}{\sqrt{k}}  - \frac{C_3\ln k}{k}  - \frac{D_3}{k}\,,
\end{align}
we assumed the following functional forms of their large-$k$ expansion \cite{korobov:13},
\begin{align}
w_{L1}(k) &\ = \frac1k\, \sum_{m = 1}^M \sum_{n = 0}^m \frac{c_{m,n} \ln^n k}{k^{m/2}}\,, \\
w_{L2,3}(k) &\ = \frac1k\,\sum_{m = 1}^M \frac{d_{m,2} \sqrt{k} + d_{m,1}\ln k + d_{m,0}}{k^{m}}\,,
\end{align}
where $c_{i,j}$ and $d_{i,j}$ are fitting coefficients. In order to ensure the stability of the
fitting, high numerical accuracy of the integrand $P_{Li}(k)$ was required, typically 10-12
significant digits.

Such accuracy turned out to be difficult to reach for the perturbed wave function part of the Breit
correction for the $2^3P$ state. The reason for this is the logarithmic singularity
\cite{korobov:13} of the perturbed wave function $\delta \phi$,
\begin{align}
\delta \phi = \frac1{(E-H)'} \,H_A^{(4)}\phi\,.
\end{align}
In order to ensure good convergence of numerical results for $\delta \phi$ we had to choose the
basis for the propagator very carefully. It was constructed as follows. We start by variationally
optimizing two symmetric second-order corrections,
\begin{align}
\delta_1E = \Big<H_A^{(4)}  \frac1{(E-H)'}  \, H_A^{(4)}\Big>\,, \\
\delta_2E = \Big<P^2  \frac1{(E-H)'} \, P^2 \Big>\,.
\end{align}
The form of $\delta_2E$ is suggested  by the expression for the leading asymptotic constant $A_1$,
Eq.~(\ref{A1}). In order to account for the logarithmic singularity present in $\delta_1E$, we
exploit the flexibility of our exponential basis functions (\ref{wf}) and emulate the singularity
by allowing the nonlinear parameters to be very large. In order to effectively span large regions
of nonlinear parameters, we used a non-uniform distribution of nonlinear parameters $\alpha_i$,
$\beta_i$, and $\gamma_i$ introduced in Ref.~\cite{pachucki:02:jpb}, typically,
\begin{equation}
\alpha_i = A_1+ \left(1/t_i^{3}-1\right)\,A_2\,,
\end{equation}
where the variable $t_i$ has a uniform quasirandom distribution over the interval $(0,1)$, and
$A_1$ and $A_2$ are the variational optimization parameters. Finally, we merge the optimized basis
sets for $\delta_1E$ and $\delta_2E$ and use the result for calculating the perturbed wave function
$\delta \phi$. Nonetheless, a large number of basis functions ($N = 3000$-$5000$) were required in order
to reach the desired accuracy.

%%%%%%%%%%%%%%%%%%%%%%%%%%%%%%%%%%%%%%%%%%%%%%%%%%%%%%%%%%%%%%%%%%%%%%%%%%%%%%%%%

\section{Results}

Our numerical results for the asymptotic expansion coefficients and the relativistic corrections to
the Bethe logarithm are presented in Table~\ref{tab:1}. For the $2^3\!S$--$2^3\!P$ transition
energy in helium, the total relativistic correction to the Bethe logarithm, $E_L = E_{L1} + E_{L2}
+ E_{L3}$, amounts to $E_L = -4.9665\,(15)$~MHz. This can be compared with the estimate of
Ref.~\cite{drake:88, pachucki:17}  obtained from the hydrogenic results by rescaling the electron density at
the origin. For the $2^3\!S$--$2^3\!P$ transition energy, this approximation yields
\begin{align} \label{ELhydr}
E_{L}({\rm appr})&\  = \alpha^7\,m\,Z^3\,\Big[{\cal L}(2s)-\frac13{\cal L}(2p_{1/2})-\frac23{\cal L}(2p_{3/2})\Big]\,
  \nonumber\\& \times
     \Big[\big< \delta^3(r_1)+\delta^3(r_2)\big>_{2^3\!S}-\big< \delta^3(r_1)+\delta^3(r_2)\big>_{2^3\!P}\Big]\,,
\end{align}
with ${\cal L}(2s) = -28.350\,965$, ${\cal L}(2p_{1/2}) = -0.795\,650$, and ${\cal L}(2p_{3/2}) =
-0.584\,517$  \cite{jentschura:05:sese,jentschura:03}. The corresponding numerical value is
$E_{L}({\rm appr}) = -3.7\,(0.9)$~MHz, where we assumed a 25\% uncertainty, like in
Ref.~\cite{pachucki:17}.

A complete treatment of the $\alpha^7m$ correction requires calculations of the two remaining terms
in Eq.~(\ref{eq1}), which will be addressed to in our future investigations. The numerical
contribution of these terms is expected to be comparable to that of $E_L$. In particular, for the
$2^3\!S$--$2^3\!P$ transition energy in helium, the hydrogenic approximation for the remaining
contribution yields $-4.3\,(1.1)$~MHz.

In summary, in this work we report calculations of the relativistic corrections to the Bethe
logarithm for the $2\,^3\!S$ and $2\,^3\!P$ states of helium. This is the first step on the path to
calculating the complete QED contribution of order $\alpha^7m$ to the triplet states of helium.
Being the most numerically demanding part, the calculation of the relativistic corrections to the
Bethe logarithm indicates the feasibility of the whole $\alpha^7\,m$ project.

%%%%%%%%%%%%%%%%%%%%%%%%%%%%%%%%%%%%%%%%%%
\begin{table}
\caption{Numerical results for the relativistic corrections to the Bethe logarithm and asymptotic
expansion constants for the $2^3\!S$ and
$2^3\!P$ (centroid) states of helium, in atomic units. \label{tab:1}
}
\begin{ruledtabular}
\begin{tabular}{l..}
\multicolumn{1}{l}{Term} & \multicolumn{1}{c}{$2^3S$} & \multicolumn{1}{c}{$2^3P$} \\
\hline\\[-5pt]
$D$      &     33.184x\,142\,629          &  31.638x\,617\,831\,(1)     \\[5pt]
$A_1$    &    -33.989x\,031\,782\,(2)   &    -31.977x\,565\,646   \\
$D_1$    &   -132.158x\,242\,69\,(5)    &   -127.498x\,493\,92\,(10)     \\
$E_{L1}$ &   -45.129x1\,(35)            &    -41.717x5\,(40)           \\[5pt]
$A_2$     &  42.692x\,780\,038          &   40.253x\,149\,916          \\
$D_2$     &  -53.768x\,709\,997\,(3)    &  -50.260x\,445\,50\,(9)     \\
$E_{L2}$  &   335.867x5\,(36)           &  319.160x1\,(36)            \\[5pt]
$G_3$     &    0.032x\,569\,625         &   -0.065x\,018\,180       \\
$F_3$     &    2.121x\,589\,807         &    2.079x\,835\,929       \\
$A_3$     &  -49.768x\,158\,799         &  -47.453x\,391\,8\,(3)     \\
$D_3$     &   1\,175.043x\,968\,722\,(4)  &  1\,121.202x\,870\,92\,(12)  \\
$E_{L3}$  &  -1\,095.043x\,9\,(3)       &  -1\,045.271x\,(8)         \\
\end{tabular}
\end{ruledtabular}
\end{table}

%%%%%%%%%%%%%%%%%%%%%%%%%%%%%%%%%%%%%%%%%%%%%%%%%%%%%%%%%%%%%%%%%%%%%%%%%%%%%%%%%

\begin{acknowledgments}
  Authors acknowledge help from M. Puchalski in numerical evaluation of asymptotic coefficients.
  This work was supported by the National Science Center (Poland) Grants No. 2012/04/A/ST2/00105
  and 2017/27/B/ST2/02459. V.A.Y. acknowledges support by the Ministry of Education and Science
  of the Russian Federation Grant No. 3.5397.2017/6.7.
	V.P. acknowledges support from the Czech Science Foundation - GA\v{C}R (Grant No. P209/18-00918S).
\end{acknowledgments}

%%%%%%%%%%%%%%%%%%%%%%%%%%%%%%%%%%%%%%%%%%%%%%%%%%%%%%%%%%%%%%%%%%%%%%%%%%%%%%%%%
\appendix

\section{Asymptotic coefficients of $\bm P_{\bm {L1}}$}
\label{sec:app1}

Here we derive the coefficients $A_1$, $B_1$, $C_1$, and $D_1$ of the large-$k$ expansion of
$P_{L1}$ given by Eq.~(\ref{eq:PL1}). There are contributions coming from both the low-energy and
the high-energy regions of the virtual photon momenta. Individually, the low- and the high-energy
parts may contain divergences, which are regularized by working in $d$ dimensions and are canceled
when both parts are added together.

\subsubsection*{Low-energy part}
The low-energy part can be derived by performing a direct large-$k$ expansion of the expression
\begin{align}
&\ \delta\, \Big< P^i\frac{k}{E-H-k}P^i \Big>\,,
\end{align}
where $\delta$ denotes the first-order perturbation of the matrix element by Breit Hamiltonian
$H^{(4)}$. The coefficient $A_1$ from Eq. (\ref{eq:PL1}) comes from the perturbation of the reference-state wave function,
\begin{align} \label{A1}
  A_1 &\ = -2\,\Big< H^{(4)}\frac{1}{(E-H)'}P^2\Big> \\
      &\ = -2\,  \Big< H_A^{(4)} \frac1{(E-H)'}\, P^2 \Big>
+ 2 \left< \bigl[ Q - \lbr Q\rbr \bigr] \, P^2 \right>\,.\nonumber
\end{align}
The low-energy part of the coefficient $D_1$ is
\begin{eqnarray}
D_1^L &=& \delta\,\Big< P^i(H-E)P^i\Big> \nonumber\\
&=& \langle H^{(4)} \frac{1}{(E-H)'}[P^i,[V,P^i]]\rangle + \frac12\,\langle [P^i,[H^{(4)},P^i]]\rangle \nonumber\\
&=& D_{1a}^L+D_{1b}^L.
\end{eqnarray}
The second-order term $D_{1a}^L$ is singular. We employ the regularized form of the Breit Hamiltonian
$H_A^{(4)}$, Eq.~(\ref{eq:HA}), in order to move singularities into first-order terms and use the
dimensional regularization in order to handle the remaining divergences. The result is
\begin{widetext}
\begin{eqnarray}
D_{1a}^L &=&
-2\,Z\,\biggl\langle H_A^{(4)}\frac{1}{(E-H)'}
\biggl(\frac{\vec{r}_1}{r_1^3}\cdot\vec{\nabla}_1+\frac{\vec{r}_2}{r_2^3}\cdot\vec{\nabla}_2\biggr)\biggr\rangle +
E^{(4)} \biggl(\biggl\langle\frac2r\biggr\rangle-4E\biggr)
\nonumber\\
&&+\,\biggl\langle \biggl[\frac{Z}{r_1}\biggr]_\epsilon\biggl(E
+\biggl[\frac{Z}{r_1}+\frac{Z}{r_2}\biggr]_\epsilon-\frac{1}{r}\biggr)^2
+\frac14\,\biggl[\frac{Z^2}{r_1^4}\biggr]_\epsilon
-\frac12\,p_1^2\frac{Z}{r_1}\,p_2^2
\nonumber\\
&&
+\biggl(2E-\frac{2}{r_2}+\biggl\langle\frac{1}{r}\biggr\rangle\biggr) \pi Z\delta^{(3)}(r_1)
+p_1^i\,\frac{Z}{r_1}\biggl(\frac{\delta^{ij}}{r}+\frac{r^ir^j}{r^3}\biggr)\,p_2^j
%\nonumber\\ &&
+\,(1\leftrightarrow2)\biggr\rangle.
\end{eqnarray}
\end{widetext}
Here, $[Z/r_1]_\epsilon$ is the $d$-dimensional form of the Coulomb potential
(for details see \cite{pachucki:06:hesinglet}). The terms
$[(Z/r_1)^3]_\epsilon$ and $[Z^2/r_1^4]_\epsilon$ contain singularities
which will be canceled when combined with corresponding terms coming from the high-energy part.
Term $D_{1b}^L$ is evaluated as
\begin{align}
D_{1b}^L &\ =
\biggl\langle
- \biggl(E+\frac{Z-1}{r_2}-\frac{p_2^2}{2}\biggr)\pi Z\,\delta^{(3)}(r_1)\nonumber\\
&\ + \frac Z2\,\vec{p_1}\,\pi\,\delta^{(3)}(r_1)\,\vec{p}_1
 + (1\leftrightarrow2)\biggr\rangle\,.
\end{align}

\subsubsection*{High-energy part}

Coefficient $B_1$ can be obtained from the forward scattering two-photon exchange diagram perturbed by the Breit
Hamiltonian. The result is
\begin{align}
B_1 &\ = \sqrt{2}\,Z^2\langle\pi\bigl[\delta^{(3)}(r_1)+\delta^{(3)}(r_2)\bigr]\rangle.
\end{align}
The Breit correction to the forward-scattering three-photon exchange diagram contains both the coefficient $C_1$ and
the high-energy part of the coefficient $D_1$,
\begin{align}
C_1 &\ = Z^3\,\langle\pi\bigl[\delta^{(3)}(r_1)+\delta^{(3)}(r_2)\bigr]\rangle\,,
\end{align}
and
\begin{align}
D_1^H &\ = Z^3\,\langle \pi \bigl[\delta^{(d)}(r_1)+\delta^{(d)}(r_2)\bigr]\rangle
\biggl(-8-\frac{1}{2\epsilon} + 9\ln 2\biggr)\,.
\end{align}
Finally, the coefficient $D_1$ is the sum of the low-energy part $D_1^L$ and the high-energy part
$D_1^H$. Making use of the identity
\begin{align}\label{A11}
 \biggl[\frac{Z^2}{r_1^4}\biggr]_\epsilon = &\ -2\biggl[\frac{Z^3}{r_1^3}\biggr]_\epsilon
+ \vec{p}_1 \frac{Z^2}{r_1^2}\,\vec{p}_1 + p_2^2\frac{Z^2}{r_1^2}
 \nonumber\\
&\ -2\biggl(E+\frac{Z}{r_2}-\frac{1}{r}\biggr)\frac{Z^2}{r_1^2},
\end{align}
we write the result as
\begin{widetext}
\begin{align}
D_1 &\ =
-2\,Z\,\biggl\langle H_A^{(4)}\frac{1}{(E-H)'}
\biggl(\frac{\vec{r}_1}{r_1^3}\cdot\vec{\nabla}_1+\frac{\vec{r}_2}{r_2^3}\cdot\vec{\nabla}_2\biggr)\biggr\rangle
 +\,E^{(4)} \biggl(\biggl\langle\frac2r\biggr\rangle-4E\biggr)
 \nonumber \\
& + \biggl\langle
 \frac Z2 \,\vec{p_1}\,\pi\,\delta^{(3)}(r_1)\,\vec{p}_1
+ \frac{Z}{r_1} (E - V)^2
- \frac{X_1}{2}
-\frac12\,p_1^2\,\frac{Z}{r_1}\,p_2^2
+p_1^i\,\frac{Z}{r_1}\biggl(\frac{\delta^{ij}}{r}+\frac{r^ir^j}{r^3}\biggr)\,p_2^j
\nonumber \\&
+\biggl[E-\frac{Z+1}{r_2}+\frac{p_2^2}{2}+\biggl\langle\frac1r\biggr\rangle  +
Z^2\bigl(-7+9\ln 2\bigr)\biggr]\,\pi Z\,\delta^{(3)}(r_1)
+(1\leftrightarrow2)\biggr\rangle\,,
\end{align}
\end{widetext}
where
\begin{align}
  \biggl\langle\biggl[\frac{Z^3}{r_1^3}\biggr]_\epsilon\biggr\rangle = &\
  \biggl\langle \frac{1}{r_1^3}\biggr\rangle + Z^3\langle\pi\delta^d(r_1)\rangle\,
  \biggl(\frac{1}{\epsilon}+2\biggr)\,,\\
\biggl\langle \frac{1}{r_1^3}\biggr\rangle =&\
\lim_{a\rightarrow 0}
\biggl\langle\frac{\Theta(r_1-a)}{r_1^3} + 4\,\pi\,\delta^3(r_1)\,
(\gamma+\ln a)\biggr\rangle\,,\\
X_1 =&\ \frac{Z^2}{r_1^2}\biggl(E-V-\frac{p_2^2}{2}\biggr) -
\frac{1}{2}\,\vec p_1\frac{Z^2}{r_1^2}\vec p_1\,.
\end{align}

\section{Asymptotic coefficients of $P_{L2}$}
\label{sec:app2}

Here we derive the coefficients $A_2$, $B_2$, $C_2$, and $D_2$ of the large-$k$ expansion of
$P_{L2}$ given by Eq.~(\ref{eq:PL2}).

\subsubsection*{Low-energy part}

First we examine contributions coming from the region of low virtual photon momenta. The
coefficient $A_2$ is the leading-order term of the direct large-$k$ expansion of Eq.~(\ref{eq:PL2:0}),
with the result
\begin{align}
  A_2 = & \,\frac12 \,\biggl< 4\,(E-V)^2 - 2 \,p_1^2\,p_2^2 + 2\,(E-V) \,\vec{p}_1\cdot \vec{p}_2
  \nonumber \\ &
  + \left(\frac{\delta^{ij}}{r} + \frac{r^ir^j}{r^3}\right)\,
  \bigl(p_1^i+p_2^i\bigr) \bigl(p_1^j+p_2^j\bigr)\biggr>\,.
\end{align}
	
The low-energy part of the coefficient $D_2$ is
\begin{align}
D_2^L &\ =\langle\phi|\delta j^i\,(H-E)\,j^i|\phi\rangle \nonumber\\
      &\ = \biggl\langle \frac{1}{2} \biggl[\delta j^i,\biggl[
			-\biggl[\frac{Z}{r_1}+\frac{Z}{r_2}\biggr]_\epsilon,j^i\biggr]\biggr]\biggr\rangle
= D_{2a}^L+D_{2b}^L.
\end{align}
Individual terms are
\begin{eqnarray}
D_{2a}^L &=& \biggl\langle \frac{1}{4}\biggl[(p_1^i\,p_1^2+p_2^i\,p_2^2),
\biggl[\biggl[\frac{Z}{r_1}+\frac{Z}{r_2}\biggr]_\epsilon,j^i\biggr]\biggr]\biggr\rangle\nonumber\\
&=& \biggl\langle - \biggl[2\,\biggl(E+\frac{Z-1}{r_2}\biggr)-p_2^2\biggr]
\,\pi Z\,\delta^{(3)}(r_1)\nonumber\\
&&-\,\frac12\,\vec{p}_1\frac{Z^2}{r_1^2}\vec{p}_1 + \biggl[\frac{Z}{r_1}\biggr]_\epsilon^2
\biggl(E+\biggl[\frac{Z}{r_1}+\frac{Z}{r_2}\bigg]_\epsilon-\frac{1}{r}\biggr)
 \nonumber \\
 && -\frac12\, p_2^2\frac{Z^2}{r_1^2}+\frac12\frac{Z\vec{r}_1}{r_1^3}\cdot\frac{\vec{r}}{r^3}
  + (1\leftrightarrow2)\biggr\rangle\,,
\end{eqnarray}
and
\begin{align}
D_{2b}^L &\ = \biggl\langle \frac{1}{4}\biggl[\biggl(\frac{\delta^{ij}}{r}+\frac{r^ir^j}{r^3}\biggr)(p_1^j+p_2^j)
    ,\biggl[\frac{Z}{r_1}+\frac{Z}{r_2},j^i\biggr]\biggr]\biggr\rangle \nonumber\\
 &\ = \biggl\langle\frac{Z}{4}\biggl(\frac{\delta^{ij}}{r}+\frac{r^ir^j}{r^3}\biggr)
     \biggl(\frac{3r_1^ir_1^j-\delta^{ij}\,r_1^2}{r_1^5}\biggr) \nonumber\\
	&\	- \frac{4\pi}{3}\frac{Z}{r_2}\,\delta^{(3)}(r_1)+(1\leftrightarrow2)\biggr\rangle\,.
\end{align}

\subsubsection*{High-energy part}

Now we turn to contributions induced by high momenta of virtual photons. The coefficient $B_2$
comes from the forward-scattering two-photon exchange perturbed by $\delta j^i$ and can be evaluated to yield
\begin{align}
  B_2 =&  -Z^2\,\sqrt{2} \, \lbr 4\pi \bigl[ \delta^3(r_1) + \delta^3(r_2)\bigr]\rbr .
  \end{align}

Similarly to the case of $P_{L1}$, the coefficient $C_2$ and the high-energy part of $D_2$ are
obtained from the forward scattering three-photon exchange with additional $\delta j^i$ and $j^i$ vertices,
\begin{align}
&\ \phi^2(0)\int\frac{d^d q_1^2}{(2\pi)^d}\frac{d^d q_2^2}{(2\pi)^d}\nonumber\\
&\ \times\biggl(\frac{-4\pi Z}{q_1^2}\biggr)\biggl(\frac{-4\pi Z}{q_2^2}\biggr)
\biggl(\frac{-4\pi Z}{(\vec{q}_1-\vec{q}_2)^2}\biggr)\,\biggl(\frac{2}{q_1^2}\biggr)\,\biggl(\frac{2}{q_2^2}\biggr)\nonumber\\
&\ \times\bigg[\frac{-\frac{1}{2}q_1^2\,(\vec{q}_1\cdot\vec{q}_2)}{\bigl(\frac{q_1^2}{2}+k\bigr)\bigl(\frac{q_2^2}{2}+k\bigr)}
-\frac{q_2^2}{\bigl(\frac{q_2^2}{2}+k\bigr)} -\frac{q_1^2}{\bigl(\frac{q_1^2}{2}+k\bigr)}\biggr]\,.
\end{align}
From this expression, we derive the following results,
\begin{align}
  C_2 =&  \frac{Z^3}{2} \, \lbr 4\pi \bigl[ \delta^3(r_1) + \delta^3(r_2)\bigr]\rbr\,,
  \end{align}	
and
\begin{equation}
D_2^H = \langle Z^3\,\pi\bigl[\delta^{(d)}(r_1)+\delta^{(d)}(r_2)\bigr]\rangle
\biggl(8-\frac{1}{\epsilon}-6\ln2\biggr)\,.
\end{equation}
The total coefficient $D_2$ is then the sum of the corresponding low-energy and high-energy parts,
\begin{eqnarray}
D_2 &=& \biggl\langle\frac{Z}{4}\biggl(\frac{\delta^{ij}}{r}+\frac{r^ir^j}{r^3}\biggr)
     \biggl(\frac{3r_1^ir_1^j-\delta^{ij}\,r_1^2}{r_1^5}\biggr)\nonumber\\
&&\ - \biggl(\frac{E}{2}+\frac{3Z-1}{6r_2}
    -Z^2\frac{5-3\ln 2}{2}-\frac{p_2^2}{4}\biggr)\,4\pi Z\,\delta^{(3)}(r_1)\nonumber\\
&&\ +X_1 +\frac Z2\frac{\vec{r}_1}{r_1^3}\cdot\frac{\vec{r}}{r^3}
+(1\leftrightarrow2)\biggr\rangle\,.
\end{eqnarray}

\section{Asymptotic coefficients of $P_{L3}$}
\label{sec:app3}

We now turn to the derivation of the coefficients $G_3$, $F_3$, $A_3$, $B_3$, $C_3$, and $D_3$ of
the large-$k$ expansion of $P_{L3}$ given by Eq.~(\ref{eq:PL3}), which is the most complicated
part.

\subsubsection*{Low-energy part}

In order to derive contributions coming from the low photon momenta, we first make a large-$k$
expansion of the propagator $1/(E-H-k)$ in Eq.~(\ref{eq:PL3:0}). In the obtained expression, we
then make a small-$k$ expansion and keep the $k^2$ contribution.

Using the angular average identity in $d$ dimensions (with $\hat{k} = \vec{k}/k$)
\begin{align}
&\ \int\frac{d\Omega_k}{4\pi} \,\hat{k}^m\hat{k}^n(\delta^{ij}-\hat{k}^i\hat{k}^j)\nonumber\\
&\ =\frac{1}{d\,(d+2)}\big[(d+1)\,\delta^{ij}\delta^{mn}-\delta^{im}\delta^{jn}-\delta^{in}\delta^{jm}\big]\,,
\end{align}
we get
\begin{align}
G_3 &\ = -\frac32\sum_{a,b=1,2}\int\frac{d\Omega_k}{4\pi}\,(\delta^{ij}-\hat{k}^i\hat{k}^j)\,
%\nonumber\\ &\ \times\,
\delta_{k^2}' \Big< p_a^i\,e^{-i\vec{k}\cdot(\vec{r}_a-\vec{r}_b)}\,p_b^j\Big>\nonumber\\
&\ = \frac1{5}\, \Bigl< p_1^i \bigl( 2\delta^{ij} r^2 - r^i r^j \bigr) p_2^j \Bigr>\,,
\end{align}
where the symbol $\delta_{k^2}'$ stands for performing a small-$k$ expansion and taking the
coefficient at the $k^2$ term.

Analogously, the next coefficient $F_3$ is obtained as
\begin{align}
F_3 =&\  \frac32\sum_{a,b=1,2}\int\frac{d\Omega_k}{4\pi}\,(\delta^{ij}-\hat{k}^i\hat{k}^j)\nonumber\\
&\ \times\,\delta_{k^2}'\Big< p_a^i\,e^{-i\vec{k}\cdot\vec{r}_a}\,
(H-E)\,p_b^j\,e^{i\vec{k}\cdot\vec{r}_b}\Big>\nonumber\\
= &\ \biggl<  E - V - \frac1{5 r} \biggr>\,.
\end{align}
Furthermore,
\begin{align}
A_3 = &\ -\frac32\sum_{a,b=1,2}\int\frac{d\Omega_k}{4\pi}\,(\delta^{ij}-\hat{k}^i\hat{k}^j)\nonumber\\
&\ \times\,\delta_{k^2}'\,\Big< p_a^i\,e^{-i\vec{k}\cdot\vec{r}_a}\,
(H-E)^2\,p_b^j\,e^{i\vec{k}\cdot\vec{r}_b}\Big>\nonumber\\
= &\
\frac{1}{10} \biggl<
Z^2\frac{4\,(\vec{r}_1\cdot\vec{r}_2)\,r^2-2\,(\vec{r}_1\cdot\vec{r})\,(\vec{r}_2\cdot\vec{r})}{r_1^3r_2^3}
-\frac{2}{r^2}
\nonumber \\ &\
- \frac45 \,(E-V)^2
 - 6 \,(\vec{p}_1\cdot\vec{p}_2)^2 + 6 \,p_1^2\,p_2^2
\nonumber \\ &\
+ \biggl[\,p_2^j\,\biggl(\frac{Zr_1^i}{r_1^3}-\frac{r^i}{r^3}\biggr)
(3\,\delta^{ik}r^j+3\,\delta^{ij}r^k-2\,\delta^{jk}r^i)\,p_2^k
\nonumber\\ &
+2\, \frac{Z\vec{r}_1}{r_1^3}\cdot\frac{\vec{r}}{r}
-2\pi Z \,\delta^3(r_1) + (1\leftrightarrow2)\biggr]
\biggr>\,.
\end{align}
The low-energy part of the coefficient $D_3$ is the most complicated term and thus will be
discussed in some detail. The starting expression is
\begin{align} \label{L3a}
D_3^L = &\ \frac32\sum_{a,b=1,2}\int\frac{d\Omega_k}{4\pi}\,(\delta^{ij}-\hat{k}^i\hat{k}^j)\nonumber\\
&\ \times\,\delta_{k^2}'\,\Big< p_a^i\,e^{-i\vec{k}\cdot\vec{r}_a}\,
(H-E)^3\,p_b^j\,e^{i\vec{k}\cdot\vec{r}_b}\Big>\,.
\end{align}
It is convenient to split the above expression into two parts, with $a=b$ and $a\neq b$. The first
part can be evaluated with help of the identity
\begin{align}
&\ e^{-i\vec{k}.\vec{r}}\,f(p)\,e^{i\vec{k}.\vec{r}}=f(p+k)\,.
\end{align}
We obtain
\begin{align}
D_{3a}^L = &\ \frac32\sum_{a=1,2}\int\frac{d\Omega_k}{4\pi}\,(\delta^{ij}-\hat{k}^i\hat{k}^j)\\
&\ \times\,\delta_{k^2}'\,\Big<p_a^i\,
\Big(H-E+\vec{p}_a\cdot\vec{k}+k^2/2\Big)^3\,p_a^j\Big>\nonumber.
\end{align}
After straightforward but tedious manipulations that involve expanding the matrix element in small
$k$ and retaining the coefficient in front of $k^2$ and using identities
\begin{align}
 &\ \biggl[\,p_1^2,\biggl[\,p_1^2,\frac{Z}{r_1}\biggr]\biggr]  =
4\biggl[\frac{Z^2}{r_1^4}\biggr]_\epsilon - 4\frac{Z\vec{r}_1\cdot\vec{r}}{r_1^3r^3},\nonumber\\
&\ \biggl[\,p_1^2,\biggl[\,p_1^2,\frac{1}{r}\biggr]\biggr]  =
2\frac{Z\vec{r}_1\cdot\vec{r}}{r_1^3r^3} - \frac{2}{r^4}
- P^i P^j\frac{3r^i r^j-\delta^{ij}r^2}{r^5},\nonumber\\
&\ p_1^i\,[\,p_1^i,[V,p_1^j]]\,p_1^j  =
\biggl[\frac{Z^2}{r_1^4}\biggr]_\epsilon - \frac32\,\frac{Z\vec{r}_1\cdot\vec{r}}{r_1^3r^3}
+ \frac{1}{2\,r^4}\nonumber\\
&\
+\frac14\,P^i P^j\frac{3r^i r^j-\delta^{ij}r^2}{r^5}
+ \frac12\,p_1^i\,[\,p_1^j,[V,p_1^j]]\,p_1^i\nonumber\\
&\ -\,\biggl(E+\frac{Z}{r_2}-\frac{1}{r}-\frac{p_2^2}{2}\biggr)\,4\pi Z\,\delta^{(3)}(r_1)\,,
\end{align}
as well as Eq.~(\ref{A11}), we arrive at
\begin{widetext}
\begin{eqnarray}
D_{3a}^L
&=& \frac{3}{2}\,
\biggl\langle \frac{7}{6}\,\frac{1}{r^4} + \frac{1}{12}\,P^i P^j\frac{3r^i r^j-\delta^{ij}r^2}{r^5}
+\frac{1}{10}\,\vec{p}_1\,4\pi Z\,\delta^{(3)}(r_1)\,\vec{p}_1
-\frac{5}{2}\frac{Z\vec{r}_1\cdot\vec{r}}{r_1^3r^3}
+\frac43\,\vec{p}_1\,\frac{Z^2}{r_1^2}\,\vec{p}_1\nonumber\\
&&-\,\frac{8}{3}\,\frac{Z^2}{r^2_1}(E-V)
+\frac43\,p_2^2\frac{Z^2}{r_1^2}
+\biggl[\frac{28}{15}\,\biggl(E+\frac{Z}{r_2}-\frac{1}{r}-\frac{p_2^2}{2}\biggr)
+Z^2\biggl(-\frac{8}{3\,\epsilon}-\frac{206}{45}\biggr)\biggr]\pi Z\,\delta^{(d)}(r_1)
\nonumber\\
&& -\,
\frac{1}{10}\,\vec{p}_1\,4\pi\,\delta^{(3)}(r)\,\vec{p}_1
+(1\leftrightarrow2)\biggr\rangle\,.\nonumber\\
\end{eqnarray}
The second term in Eq.~(\ref{L3a}) with $a\neq b$ is evaluated as
\begin{align}\label{C12}	
D_{3b}^L &\ =  \frac32\int\frac{d\Omega_k}{4\pi}\,(\delta^{ij}-\hat{k}^i\hat{k}^j)
\,\delta_{k^2}'\Big<p_1^i\,e^{-i\vec{k}\cdot\vec{r}_1}\,
(H-E)^3\,p_2^j\,e^{i\vec{k}\cdot\vec{r}_2}\Big>
+ (1\leftrightarrow2)\nonumber\\
        &\ =  \frac34\int\frac{d\Omega_k}{4\pi}\,(\delta^{ij}-\hat{k}^i\hat{k}^j)
 \,\delta_{k^2}'\Big< \big[\big[ p_1^i\,e^{-i\vec{k}\cdot\vec{r}_1},
H-E\bigr],\bigl[H-E,\bigl[H-E,\,p_2^j\,e^{i\vec{k}\cdot\vec{r}_2}\bigr]\bigr]\bigr]\Big>
  + (1\leftrightarrow2)
 =\frac34 \sum_{m=1..8} T_m\,.
\end{align}
\end{widetext}

The individual terms $T_i$ are calculated as follows:
\begin{align}
T_1 &\ = 	\int\frac{d\Omega_k}{4\pi}\,(\delta^{ij}-\hat{k}^i\hat{k}^j)\nonumber\\
&\ \times \delta_{k^2}'\Big<\big[\big[p_1^i\,e^{-i\vec{k}\cdot\vec{r}_1},
V\big],\big[V,\big[V,\,p_2^j\,e^{i\vec{k}\cdot\vec{r}_2}\big]\big]\big]\Big>\nonumber\\
&\ +(1\leftrightarrow2) = 0\,,
\end{align}
\begin{align}
T_2 &\ = 	\int\frac{d\Omega_k}{4\pi}\,(\delta^{ij}-\hat{k}^i\hat{k}^j)\nonumber\\
&\ \times \delta_{k^2}'\Big<\Big[\Big[p_1^i\,e^{-i\vec{k}\cdot\vec{r}_1},
V\Big],\Big[V,\Big[\frac{p_2^2}{2},\,p_2^j\,e^{i\vec{k}\cdot\vec{r}_2}\Big]\Big]\Big]\Big>
+(1\leftrightarrow2)\nonumber\\
&\ = \frac{2}{15}\biggl[\biggl(\frac{Z \vec{r}_1}{r_1^3}-\frac{Z \vec{r}_2}{r_2^3}\biggr)\cdot\frac{\vec{r}}{r^3}
-\frac{2}{r^4}\biggr]\,,
\end{align}
\begin{align}
T_3 &\ = 	\int\frac{d\Omega_k}{4\pi}\,(\delta^{ij}-\hat{k}^i\hat{k}^j)\nonumber\\
&\ \times \delta_{k^2}'\Big<\Big[\Big[p_1^i\,e^{-i\vec{k}\cdot\vec{r}_1},
\frac{p_1^2}{2}\Big],\Big[V,\Big[V,\,p_2^j\,e^{i\vec{k}\cdot\vec{r}_2}\Big]\Big]\Big]\Big>
+(1\leftrightarrow2)\nonumber\\
&\ = 0\,,
\end{align}
\begin{align}
T_4 &\ = 	\int\frac{d\Omega_k}{4\pi}\,(\delta^{ij}-\hat{k}^i\hat{k}^j)\nonumber\\
&\ \times \delta_{k^2}'\Big<\Big[\Big[p_1^i\,e^{-i\vec{k}\cdot\vec{r}_1},
V\Big],\Big[\frac{p_1^2}{2}+\frac{p_2^2}{2},\Big[V,\,p_2^j\,e^{i\vec{k}\cdot\vec{r}_2}\Big]\Big]\Big]\Big>
    \nonumber\\
&\
+(1\leftrightarrow2)\nonumber\\
&\ =
\frac{4}{15}
\biggl(\frac{2}{r^4}+\biggl[2Z\biggl(\frac{\delta^{ij}}{r_1^3} - 3\frac{r_1^i r_1^j}{r_1^5 }
\biggr)\frac{r^i r^j}{r^3}
+3\,\frac{Z\vec{r}_1}{r_1^3}\cdot\frac{\vec{r}}{r^3}\nonumber\\
&\ -\frac{1}{3r_2}4\pi Z\,\delta^{(3)}(r_1)+(1\leftrightarrow2)\biggr]\biggr)\,,
\end{align}
\begin{align}
T_5 &\ = 	\int\frac{d\Omega_k}{4\pi}\,(\delta^{ij}-\hat{k}^i\hat{k}^j)\nonumber\\
&\ \times\delta_{k^2}' \Big<\Big[\Big[p_1^i\,e^{-i\vec{k}\cdot\vec{r}_1},
\frac{p_1^2}{2}\Big],\Big[V,\Big[\frac{p_2^2}{2},\,p_2^j\,e^{i\vec{k}\cdot\vec{r}_2}\Big]\Big]\Big]\Big>
+(1\leftrightarrow2)\nonumber\\
&\ =
-\frac{16}{45}\,\vec{p}\,4\pi\,\delta^{(3)}(r)\,\vec{p}
+ \frac{P^iP^j}{15} \bigg(\frac{\delta^{ij}}{r^3}-3\frac{r^i r^j}{r^5}\biggr)\nonumber\\
&\ - \frac{4}{15}\,p^i\biggl(\frac{\delta^{ij}}{r^3}-3\frac{r^i r^j}{r^5}\biggr)\,p^j\,,
\end{align}
where $p^i=\frac12(p_1^i-p_2^i)$,
\begin{align}
T_6 &\ = 	\int\frac{d\Omega_k}{4\pi}\,(\delta^{ij}-\hat{k}^i\hat{k}^j)\nonumber\\
&\ \times \delta_{k^2}'\Big<\Big[\Big[p_1^i\,e^{-i\vec{k}\cdot\vec{r}_1},
V\Big],\Big[\frac{p_2^2}{2},\Big[\frac{p_2^2}{2},\,p_2^j\,e^{i\vec{k}\cdot\vec{r}_2}\Big]\Big]\Big]\Big>
+(1\leftrightarrow2)\nonumber\\
&\ = -\frac{4}{45}\,\vec{p}\,4\pi\delta^{(3)}(r)\,\vec{p}
+\frac{2P^i P^j}{15}\,\biggl(\frac{\delta^{ij}}{r^3}-3\frac{r^i r^j}{r^5}\biggr)\nonumber\\
&\ +\frac{8}{15}\,p^i\biggl(\frac{\delta^{ij}}{r^3}-3\frac{r^i r^j}{r^5}\biggr)p^j\,,
\end{align}
\begin{align}
T_7 &\ = 	\int\frac{d\Omega_k}{4\pi}\,(\delta^{ij}-\hat{k}^i\hat{k}^j)\nonumber\\
&\ \times \delta_{k^2}'\Big<\Big[\Big[p_1^i\,e^{-i\vec{k}\cdot\vec{r}_1},
\frac{p_1^2}{2}\Big],\Big[\frac{p_1^2+p_2^2}{2},\Big[V,\,p_2^j\,e^{i\vec{k}\cdot\vec{r}_2}\Big]\Big]\Big]\Big>\nonumber\\
&\ +(1\leftrightarrow2)\nonumber\\
&\ =
-\frac{8}{45}\,\vec{p}\,4\pi\delta^{(3)}(r)\,\vec{p}
+\frac{2P^i P^j}{15}\,\biggl(\frac{\delta^{ij}}{r^3}-3\frac{r^i r^j}{r^5}\biggr)\nonumber\\
&\ +\,\frac{16}{15}\,p^i\biggl(\frac{\delta^{ij}}{r^3}-3\frac{r^i r^j}{r^5}\biggr)p^j\,,
\end{align}
and
\begin{align}
T_8 &\ = 	\int\frac{d\Omega_k}{4\pi}\,(\delta^{ij}-\hat{k}^i\hat{k}^j)\nonumber\\
&\ \times \delta_{k^2}'\Big<\Big[\Big[p_1^i\,e^{-i\vec{k}\cdot\vec{r}_1},
\frac{p_1^2}{2}\Big],\Big[\frac{p_1^2+p_2^2}{2},\Big[\frac{p_2^2}{2},\,p_2^j\,e^{i\vec{k}\cdot\vec{r}_2}\Big]\Big]\Big]
\Big>\nonumber\\
&\ +(1\leftrightarrow2) = 0\,.
\end{align}
Substituting the terms $T_m$ into Eq.~(\ref{C12}), we obtain the result for the low-energy part of
the coefficient $D_3$,
\begin{widetext}
\begin{eqnarray}
D_3^L &=&
3\,\biggl\langle
\frac{1}{3}\,p^i\,\biggl(\frac{\delta^{ij}}{r^3}-3\frac{r^i r^j}{r^5}\biggr)\,p^j
-\frac{23}{90}\,\vec{p}\,4\pi\,\delta^{(3)}(r)\,\vec{p}
+ \frac{37}{30}\,\frac{1}{r^4} + \frac12\biggl\{
-\frac83\,\frac{Z^2}{r_1^2}(E-V)
+\frac43\,\vec{p}_1\,\frac{Z^2}{r_1^2}\,\vec{p}_1\nonumber\\
&&
+\,\frac{1}{10}\,\vec{p}_1\,4\pi Z\,\delta^{(3)}(r_1)\,\vec{p}_1
-\frac{61}{30}\frac{Z\vec{r}_1\cdot\vec{r}}{r_1^3 r^3}
+\frac43\,p_2^2\frac{Z^2}{r_1^2}
+\frac{4}{15}Z\,\biggl(\frac{\delta^{ij}}{r_1^3} - 3\frac{r_1^i r_1^j}{r_1^5 }\biggr)\frac{r^i r^j}{r^3}\nonumber\\
&&+\,\biggl[\frac{7}{15}\biggl(E+\frac{Z}{r_2}-\frac{p_2^2}{2}\biggr)
-\frac{23}{45}\frac{1}{r_2}-Z^2 \biggl(\frac{2}{3\,\epsilon}+\frac{103}{90}\biggr)
\biggr]\,4\pi Z\,\delta^{(d)}(r_1)+(1\leftrightarrow2)\biggr\}\biggr\rangle\,.
\end{eqnarray}
\end{widetext}

\subsubsection*{High-energy part}

The coefficient $B_3$ comes from the corresponding forward-scattering two-photon exchange diagram,
\begin{align}
  B_3 =&\ \frac{3k\sqrt{k}}{2}\int\frac{d\Omega_k}{4\pi}\,\delta_{k^2} \,\phi^2(0)\,\int\frac{d^3p}{(2\pi)^3}
	\biggl[\frac{-4\pi\,Z\alpha}{p^2}\biggr]^2\,\biggl(\frac{2}{p^2}\biggr)^2\nonumber\\
&\ \times p^i\,\frac{(-2)}{(p+k)^2+2\omega}\,p^j\,(\delta^{ij}-\hat{k}^i\hat{k}^j)\nonumber\\
&\ =
	2\,\sqrt{2}\,Z^2 \,\lbr 4\pi \bigl[ \delta^3(r_1) + \delta^3(r_2)\bigr]\rbr\,.
  \end{align}
	Similarly, the analogous forward-scattering three-photon exchange diagram gives rise to the coefficient $C_3$,
\begin{align}
  C_3 =&\ -2\, Z^3 \, \lbr 4\pi \bigl[ \delta^3(r_1) + \delta^3(r_2)\bigr]\rbr\,,
  \end{align}	
and the high-energy part of the coefficient $D_3$,
\begin{align}
D_3^H &\ =
Z^3\,\lbr 4\pi \bigl[ \delta^{(d)}(r_1) + \delta^{(d)}(r_2)\bigr]\rbr\nonumber\\
&\ \times\,\biggl(-\frac{73}{45}+\frac{2}{3\,\epsilon}+\frac{4}{3}\ln2\biggr)\,.
\end{align}	
The total coefficient $D_3$ is obtained as a sum of the low-energy part $D_3^L$ and the high-energy
part $D_3^H$. Using the identity,
\begin{align}
&\ p^i\,\biggl(\frac{\delta^{ij}}{r^3}-3\frac{r^i r^j}{r^5}\biggr)\,p^j
 =\frac{2\pi}{3}\,\vec{p}\,\delta^{(3)}(r)\,\vec{p} -\frac{1}{2\,r^4}\nonumber\\
&\ +\frac14\,\biggl(\frac{Z\vec{r}_1}{r_1^3}-\frac{Z\vec{r}_2}{r_2^3}\biggr)
\cdot\frac{\vec{r}}{r^3}\,,
\end{align}
we express the final result for $D_3$ as
\begin{widetext}
  \begin{align}
D_3 =&\ \frac{1}{5}
\biggl< -3\,\vec{p}\,\,4\pi\,\delta^{(3)}(r)\,\vec{p}
+ \frac{16}{r^4}
+\biggl[2\,Z\,\biggl(\frac{\delta^{ij}}{r_1^3} -
  3\frac{r_1^i r_1^j}{r_1^5 }\biggr)\frac{r^i r^j}{r^3} - 20\,X_1
+\frac{3}{4}\,\vec{p}_1\,4\pi Z\,\delta^{(3)}(r_1)\,\vec{p}_1
-14\,\frac{Z\vec{r}_1}{r_1^3}\cdot\frac{\vec{r}}{r^3}\nonumber\\
&\ +\biggl(\frac{7E}{2}+\frac{7Z}{2}\frac{1}{r_2} -\frac{23}{6}\frac{1}{r_2}- \frac{7}{4}\,p_2^2
-\frac{83Z^2}{4}+10\,Z^2\ln2\biggr)\,4\pi Z\,\delta^{(3)}(r_1)
+(1\leftrightarrow2)\biggr]\biggr> \,.
  \end{align}
\end{widetext}
The operator $\vec{p}_1\,4\pi Z\,\delta^{(3)}(r_1)\,\vec{p}_1$ requires some clarifications,
because its expectation value is conditionally converging.
It should be calculated with the implicit projection into the $L=0$ state between $\vec{p}_1$ operators
and this requirement comes from the dimensional regularization.

\section{Wave functions in Cartesian coordinates}

\label{app:wf}

Following Schwartz \cite{schwartz:61}, we use the following representations of the wave functions,
with $F  \equiv F(r_1,r_2,r)$, $G  \equiv G(r_1,r_2,r)$, and the upper sign corresponding to the
singlet function and the lower sign, to the triplet function,
\begin{align}
\phi\left(^{1,3}\!S^e\right) = F \pm (1\leftrightarrow2)\,,
\end{align}
\begin{align}
\vec{\phi}\left(^{1,3}\!P^o\right) &\ = \vec{r}_1 \, F \pm (1\leftrightarrow2)\,, \\
\vec{\phi}\left(^{1,3}\!P^e\right) &\ = \vec{r}_1 \times \vec{r}_2 \, F \pm (1\leftrightarrow2)\,,
\end{align}
\begin{align}
\phi^{ij}\left(^{1,3}\!D^o\right) &\ = \Bigl( \epsilon^{iab}r_1^a r_2^b r_1^j + \epsilon^{jab}r_1^a r_2^b r_1^i\Bigr) F
\pm (1\leftrightarrow2)\,,
\end{align}
\begin{align}
\phi^{ij}\left(^{1,3}\!D^e\right) &\ = \left(r_1^ir_1^j - \frac13\,\delta^{ij}r_1^2\right) F
 \nonumber \\ &
+ \frac12 \left(r_1^ir_2^j + r_2^ir_1^j - \frac23\, \delta^{ij}\,\vec{r}_1\cdot \vec{r}_2\right) G  \pm (1\leftrightarrow2)\,,
\end{align}
and
\begin{align}
\phi^{ijk}\left(^{1,3}\!F^o\right) &\ =
\Big[r_1^ir_1^jr_1^k - \frac15\,r_1^2\left(\delta^{ij}r_1^k+ \delta^{ik}r_1^j+ \delta^{jk}r_1^i
\right) \Big] F
 \nonumber \\ &
+ \frac13 \, \Bigl[ r_1^i r_1^j r_2^k + r_1^i r_2^j r_1^k + r_2^i r_1^j r_1^k
 \nonumber \\ &
- \frac15 \,\delta^{ij} \left( r_1^2 r_2^k + 2\, \vec{r}_1\cdot\vec{r}_2 r_1^k\right)
 \nonumber \\ &
- \frac15 \,\delta^{ik} \left( r_1^2 r_2^j + 2\, \vec{r}_1\cdot\vec{r}_2 r_1^j\right)
 \nonumber \\ &
- \frac15 \,\delta^{jk} \left( r_1^2 r_2^i + 2\, \vec{r}_1\cdot\vec{r}_2 r_1^i\right) \Bigr] G
\pm (1\leftrightarrow2)\,.
\end{align}

\end{document}